\documentclass[12pt,epsf]{article}
\usepackage{graphicx, amsmath}

\begin{document}
\sloppy
\begin{flushright}{SIT-HEP/TM-49}
\end{flushright}
\vskip 1.5 truecm
\centerline{\large{\bf Cosmological perturbations from an
inhomogeneous}}
\centerline{\large{\bf phase transition}}
\vskip .75 truecm

\centerline{\bf Tomohiro Matsuda\footnote{matsuda@sit.ac.jp}}
\vskip .4 truecm
\centerline {\it Laboratory of Physics, Saitama Institute of Technology,}
\centerline {\it Fusaiji, Okabe-machi, Saitama 369-0293, 
Japan}
\vskip 1. truecm

\makeatletter
\@addtoreset{equation}{section}
\def\theequation{\thesection.\arabic{equation}}
\makeatother
\vskip 1. truecm

\begin{abstract}
\hspace*{\parindent}
A mechanism for generating metric perturbations in inflationary models
 is considered. 
Long-wavelength inhomogeneities of light scalar fields
 in a decoupled sector may give rise to superhorizon fluctuations of
 couplings and masses in the low-energy effective action. 
Cosmological phase transitions may then occur that are not simultaneous
 in space, but occur with time lags in different Hubble patches that
 arise from the long-wavelength inhomogeneities. 
Here an interesting model in which cosmological perturbations may be
 created at the electroweak phase transition is considered. 
The results show that phase transitions may be a generic source of
 non-Gaussianity.   
\end{abstract}

\newpage
\section{Introduction}
The energy density during inflation is dominated by the inflaton
potential energy. 
At the end of inflation, the energy stored in the inflaton potential is
converted into particles, which decay and reheat the Universe by
thermalization to start the standard hot big bang phase. 

In this paper we consider phase transition from a phase $A$ to another
phase $B$, which are distinguished by the scaling of the energy density
of a component $\rho_i$.
Namely, we consider a scaling of the energy density  $\rho_i\propto
a^{-n_A}$ in phase $A$ and $\rho_i\propto a^{-n_B}$ in phase $B$, where
$n_A \ne n_B$ causes generation of the density perturbations when the
phase transition is inhomogeneous in space.
The mechanism is very general and can be applied to many other models in
which the transition between phases of different scaling is 
inhomogeneous in space, even if the transition is not a ``phases
transition'' in the strict meaning.

Before discussing inhomogeneous phase transition, we review the
mechanism of inhomogeneous reheating \cite{Dvali-Mass-dom}
to illustrate the basis of inhomogeneous scenarios.  
In Ref.\cite{Dvali-Mass-dom}, it has been argued that in realistic
models of inflation the coupling of the inflaton to matter can be
 determined by the vacuum expectation values of fields in the underlying
 theory. 
If those fields (in the string theory they would be moduli fields from the
compactified space) are light during inflation, 
they will fluctuate leading to density perturbations through the 
inhomogeneities of the coupling constants.

\begin{figure}[h]
 \begin{center}
\begin{picture}(440,250)(0,0)
\resizebox{18cm}{!}{\includegraphics{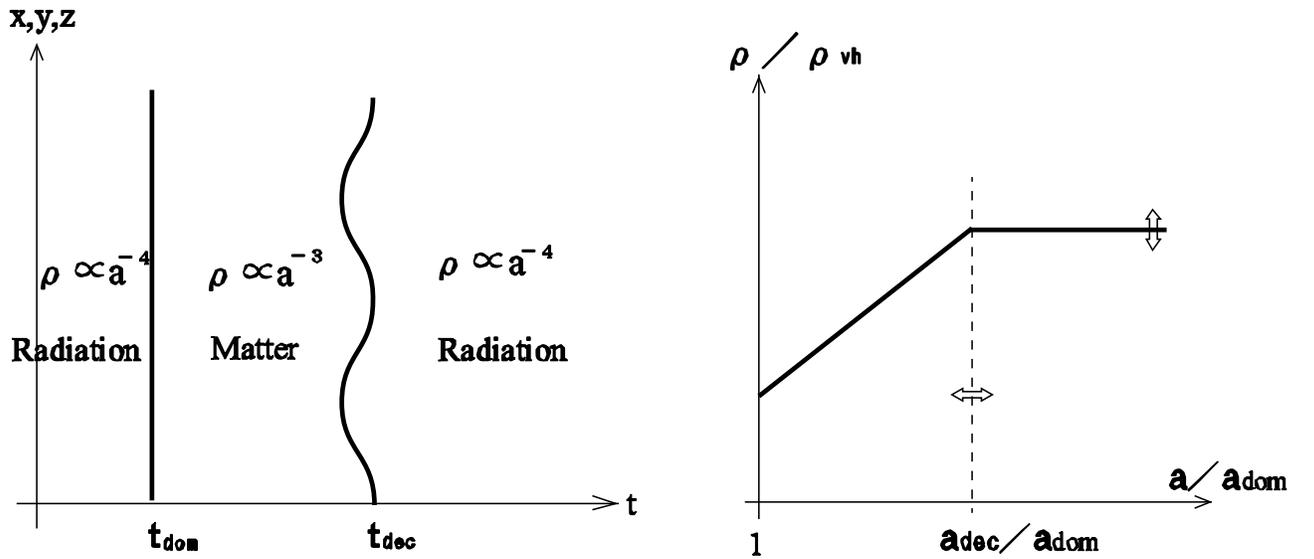}} 
\end{picture}
\label{fig:mass-dom}
 \caption{Due to the $\Gamma_\psi$ inhomogeneity, the decay of the $\psi$
  particles does not occur simultaneously in space, which leads to a
  fluctuation of $t_{dec}$ and $\rho_{dec}/\rho^r_{vh}$.
Thus, the evolution of the energy density is different in different
  patches, which results in density fluctuations.}
 \end{center}
\end{figure}

If the density perturbations created during inflation are negligible and
the Universe after inflation is filled with particles $\psi$ of mass
$M_{\psi}$ and decay rate $\Gamma_\psi<H_I$, where $H_I$ is the Hubble
parameter during inflation, spatial inhomogeneities in $\Gamma_\psi$ may
lead to density perturbations when the particles decay into radiation. 
In deriving the magnitude of the density perturbations arising from the
inhomogeneity, it is useful to compare the energy density in a region to
the virtual hidden radiation $\rho_{vh}^r$, which scales as\footnote{The
``virtual hidden radiation'' is introduced just to keep track of the
unperturbed spatially flat hypersurfaces.}
\begin{equation}
\label{rad-rho}
\rho^r_{vh}\propto a^{-4},
\end{equation}
and calculate the density perturbations on a uniform $\rho^r_{vh}$
surface. 
Here, we assume that there is no energy transition between the radiation
density $\rho^r_{vh}$ and other components of the Universe. 
Assuming that the domination by $\psi$ particles starts at 
$a_{dom}\equiv a(t_{dom})$ when $\rho_{dom}\equiv
\rho(t_{dom})\simeq\rho_\psi(t_{dom})\simeq M_\psi^4$,\footnote{Here we
assume that the mass of the $\psi$ 
particles is a constant. Unlike the original argument in
ref.\cite{Dvali-Mass-dom},  we consider a uniform $\rho_{dom}$ 
and $\delta M_\psi=0$ to simplify the argument.} 
the energy of the $\psi$ particles scales as matter in the domination
interval $t_{dom}<t<t_{dec}$:
\begin{equation}
\rho_{\psi}\simeq \rho \propto a^{-3},
\end{equation}
with decay time $t=t_{dec}$  defined by
\begin{equation}
\rho_{dec}\equiv \rho(t_{dec})\simeq\Gamma_\psi^2M_p^2.
\end{equation}
Outside the domination interval, we assume that the energy density
scales as radiation. 
Fig.(\ref{fig:mass-dom}) shows a schematic representation of the
inhomogeneous boundary that creates density fluctuations. 
Note that in this model the delay of the $\psi$-decay causes a
delay in the evolution of the energy density.
The calculation of the density perturbation is straightforward. 
Considering the $\psi$ domination interval, we find
\begin{equation}
\label{inste}
\left(\frac{a_{dec}}{a_{dom}}\right)^3=\frac{\rho_{dom}}{\rho_{dec}}
=\frac{\rho_{dom}}{\Gamma_\psi^2 M_p^2},
\end{equation}
where $\rho_{dom}$ and $M_p$ are uniform in space,
while $\Gamma_\psi$ is inhomogeneous.
Using $\rho^r_{vh}$ in Eq.(\ref{rad-rho}),
we can find the energy density after the decay:
\begin{equation}
\label{inste2}
\rho\propto \frac{a_{dec}}{a_{dom}}\rho^r_{vh}=
\frac{\rho_{dom}^{1/3}}{\Gamma_\psi^{2/3} M_p^{2/3}}\rho^r_{vh},
\end{equation}
where the ratio $\rho/\rho^r_{vh}$ is a time-independent constant after
the decay. 
The density perturbation on a uniform $\rho^r_{vh}$ surface is
thus given by 
\begin{equation}
\frac{\delta \rho}{\rho}=-\frac{2}{3}
\frac{\delta \Gamma_\psi}{\Gamma_\psi}, 
\end{equation}
which reproduces the limit $\Gamma_\psi/H_I\rightarrow 0$ in
Ref.\cite{IH-R}. 

Another way to generate cosmological perturbations from an inhomogeneous
boundary is to consider an inhomogeneous end for the inflationary phase
\cite{End-Modulated,End-multi, End-trap, End-string}.
For inflationary expansion, the equation for the number of
e-foldings is 
\begin{equation}
N \equiv \ln \frac{a(t_e)}{a(t_N)},
\end{equation}
where $t_N$ is the time when the long-wavelength inhomogeneity exits the
horizon and $t_e$ is the time when inflation ends.
We define $\phi_N\equiv \phi(t_N)$ and $\phi_e\equiv\phi(t_e)$ for the
inflaton field $\phi$. 
Using $\rho^r_{vh}$ and repeating the calculation given above, in
place of Eq. (\ref{inste}) and Eq. (\ref{inste2}), we obtain 
\begin{equation}
\left(\frac{a(t_e)}{a(t_N)}\right)^0=
\frac{\rho(t_e)}{\rho(t_N)}
\end{equation}
and
\begin{equation}
\rho\propto \left(\frac{a(t_e)}{a(t_N)}\right)^4\rho^r_{vh}=
e^{4N}\rho^r_{vh}.
\end{equation}
If we assume instant decay and instant thermalization after inflation,
the energy density of the Universe after inflation scales as radiation
and $\rho/\rho^r_{vh}$  is a time-independent constant after inflation. 
Therefore, the density perturbation on a uniform $\rho^r_{vh}$ 
surface, which is caused by the inhomogeneities in $N$, is given by 
\begin{equation}
\frac{\delta \rho}{\rho}=4\delta N,
\end{equation}
and recovers the conventional delta-N formula $\zeta =\delta
N$.\footnote{See Appendix A for the definition of the curvature 
perturbation $\zeta$ and the $\delta
N$ formula that relates $\zeta$ to $\delta N$.}
More specifically, we can calculate $\delta N$ from the $\phi_e$
inhomogeneity in the inflationary scenario using a very simple equation
$\delta N_{end} \simeq (\partial N/\partial \phi_e)\delta \phi_e$.
In most inflationary scenarios, $N$ is given explicitly by
$\phi_N$ and $\phi_e$.

Considering the two scenarios discussed above, the curvature
perturbations created by the inhomogeneous boundaries are natural
consequences of the inhomogeneities arising from long-wavelength
fluctuations of light fields. 
In this paper, we consider inhomogeneous
phase transitions in which the critical temperature is not homogeneous
in space. 
If the potential energy dominates during a short interval, the
phase may be dubbed mini-inflation. 
Following the uniform $\rho_{vh}$ calculation discussed above, 
we can calculate the density perturbations created at the phase
boundary. 
To calculate the density perturbations we assume (1) the
beginning of the phase occurs simultaneously in space, but the end is
inhomogeneous, (2) the transition occurs instantly just after the
interval and (3) all the energy stored in the potential is translated
into radiation. 
Complementary scenarios for more generic situations
require numerical study and are highly model-dependent, thus they will
be considered in future works. 
However, we consider a particularly attractive model, featuring the
possibility of inhomogeneous phase 
transitions at the electroweak (or more generically, unification) scale
that may lead to the creation of a significant level of
non-Gaussianity. 

\section{The model}

\subsection{Simple model for second order phase transition}
To illustrate some typical features of finite temperature effects, we
consider a real scalar field and a potential: 
\begin{eqnarray}
{\cal L}&=&\frac{1}{2}\partial_{\mu}\phi\partial^{\mu}\phi-V(\phi)\nonumber\\
V(\phi)&=&V_0-\frac{1}{2}m_\phi^2\phi^2+\frac{1}{4}\lambda\phi^4,
\end{eqnarray}
where $V_0$ is tuned so that the cosmological constant vanishes at the
true minimum. 
The phenomenon of high-temperature symmetry restoration
can be understood by the finite-temperature effective potential given by
\cite{EU-book-kolb}
\begin{equation}
V_T(\phi_c)=V(\phi_c)+\frac{T^4}{2\pi^2}\int^{\infty}_{0}
dx \ln \left[1-\exp\left(
-\sqrt{x^2+\frac{-m_\phi^2+3\lambda\phi_c^2}{T^2}}
\right)
\right],
\end{equation}
where $V(\phi_c)$ is the one-loop potential for zero-temperature with
the classical field $\phi_c$:
\begin{equation}
V(\phi_c)=-\frac{1}{2}m_\phi^2\phi_c^2+\frac{1}{4}\lambda\phi_c^4
+\frac{1}{64\pi^2}\left(-m_\phi^2+3\lambda\phi_c^2\right)^2
\ln \left(\frac{-m_\phi^2+3\lambda\phi_c^2}{\mu^2}\right),
\end{equation}
where $\mu$ is a renormalization mass scale.
At high temperatures, $V_T$ can be expanded as
\begin{equation}
V_T\simeq V(\phi_c)+\frac{1}{8}\lambda T^2 \phi_c^2+ {\cal O}(T^4),
\end{equation}
which suggests that the temperature-corrected effective mass at 
$\phi_c=0$ changes sign at a critical temperature 
\begin{equation}
T_c \simeq \frac{2m_\phi}{\lambda^{1/2}}.
\end{equation}
In a more general situation, one may introduce couplings to the fields in
the background thermal bath.
If the couplings of $\phi$ to the fields in the background thermal
bath are more significant than the self-coupling, a typical form of the
potential with a thermal correction term is given by
\begin{equation}
V=V_0+\left(g^2 T^2 -\frac{1}{2}m_\phi^2\right)\phi^2+ ...,
\end{equation}
where $g$ denotes the effective coupling of $\phi$ to the fields in the
thermal bath. 
In this case, the critical temperature is given by
\begin{equation}
T_c \simeq \frac{m_\phi}{2g}.
\end{equation}
In this section, we consider the latter case where $T_c$ is given by
$T_c \simeq \frac{m_\phi}{2g}$. 

\begin{figure}[h]
 \begin{center}
\begin{picture}(400,270)(0,0)
\resizebox{12cm}{!}{\includegraphics{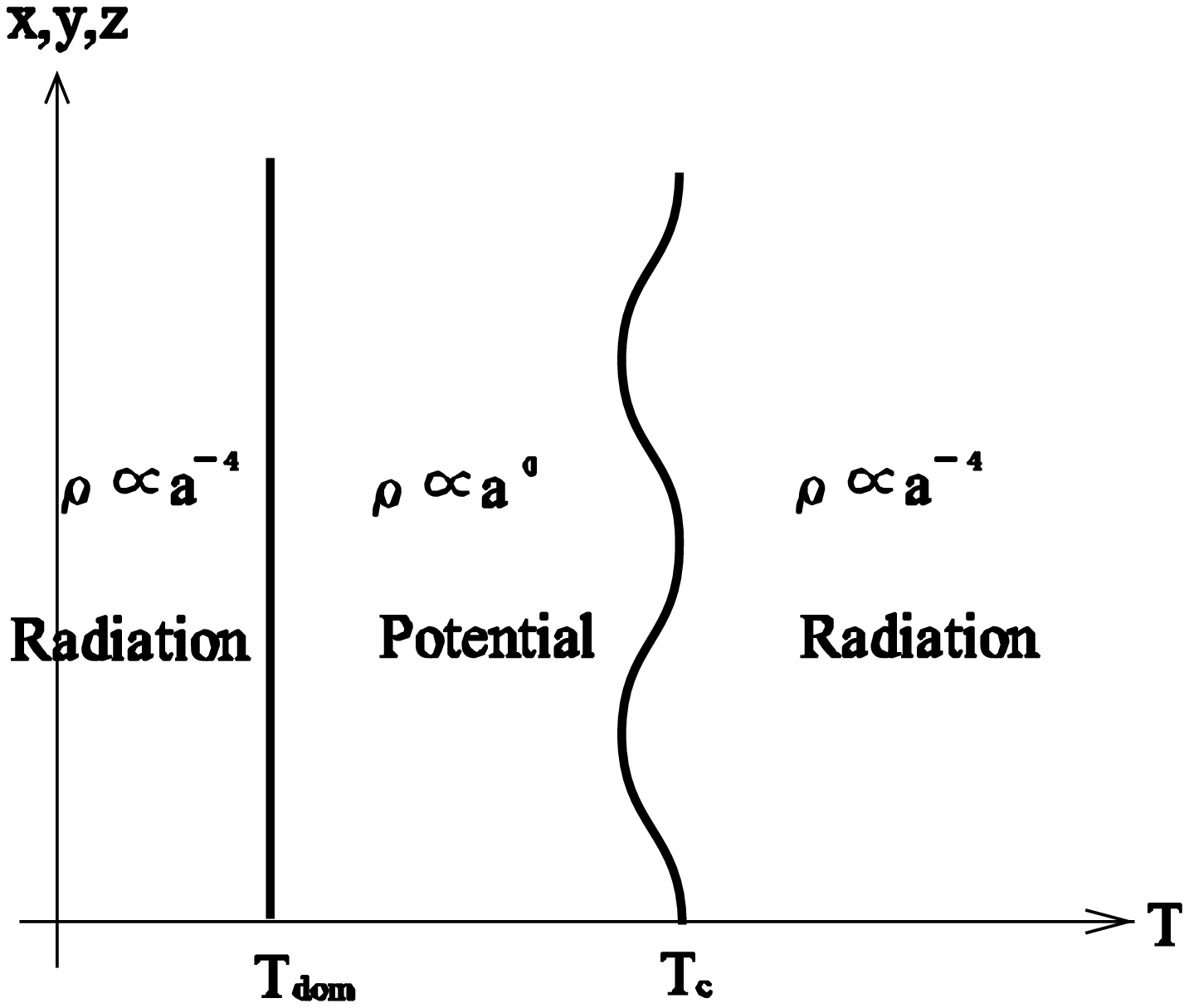}} 
\end{picture}
\caption{Initially, the Universe is dominated by radiation. The
  potential energy then starts to dominate at $T=T_{dom}$. 
The domination by the potential ends at $T=T_c$, where a phase transition
  occurs. Radiation domination starts after the phase transition.} 
\label{fig:case1}
 \end{center}
\end{figure}

The phase transition is second order in the model discussed above. 
We consider two distinct cases: 
\begin{enumerate}
\item The energy density of the Universe is dominated by the potential
      energy $V_0$ during the interval $T_{dom}>T>T_c$. 
      The Universe is then dominated by
      radiation, due to instant decay. We assume that all the energy
      stored in the potential is converted into radiation just after the
      phase transition (i.e., we assume $T_c=T_{dec}$. See also 
      Fig.\ref{fig:case1}). 
\item The energy density of the Universe is still dominated by
      radiation at $T=T_c$. After the phase transition at $T=T_c$, all the
      potential energy is converted into non-relativistic particles
      $\psi$ that scale as matter. 
      The interval of the radiation domination may
      end at $T=T_{dom}$ when $\rho_{\psi}/\rho_{rad}\simeq1$, or more
      generically the $\psi$ particles may decay into radiation at
      $T=T_{dec}$ before
      the domination. In this scenario, the inhomogeneous phase
      transition causes the inhomogeneities of
      the matter density. 
      See also Ref.\cite{IH-PR-withoutID} in which
      the inhomogeneities of the curvatons are generated by
      inhomogeneous preheating.
\end{enumerate}

In the former case, calculating the density perturbation is
straightforward. 
We assume that the interval of domination by the
potential energy starts at $T=T_{dom}\equiv T(t_{dom})$.
Considering $\rho_{vh}^r\propto a^{-4}$ as before, after the phase
transition at $T=T_{c}\equiv T(t_c)$ we find that 
\begin{equation}
\rho_\phi(t)\propto \left(\frac{T_{dom}}{T_c}\right)^4\rho_{vh}^r.
\end{equation}
Just after the phase transition, the potential energy is converted into
radiation. 
The energy density perturbation on a uniform $\rho_{vh}^r$ surface is
thus given by 
\begin{equation}
\frac{\delta \rho}{\rho}=-4\frac{\delta T_c}{T_c}
=-4\frac{\delta m_\phi}{m_\phi}+4\frac{\delta g}{g},
\end{equation}
where the curvature perturbation is given by 
\begin{equation}
\label{delm}
\zeta =\frac{1}{4}\frac{\delta \rho}{\rho} =-\frac{\delta T_c}{T_c}
=-\frac{\delta m_\phi}{m_\phi}+\frac{\delta g}{g}.
\end{equation}
This result can be obtained alternatively from the delta-N formula 
$\zeta =\delta N$.
For inflationary expansion during the $V_0$-dominated
interval\cite{EU-book}, the number of e-foldings is given by
\begin{equation}
N=\ln \left(\frac{T_{dom}}{T_c}\right),
\end{equation}
which leads to $\zeta =\delta N = -\delta T_c/T_c$.
In order to calculate the pure contribution from the inhomogeneous phase
transition, we assume that all the energy stored in the potential is
converted into radiation just after inflation. 
To understand the light-field potential, we consider a specific choice
for the $\sigma$-dependent mass: 
\begin{equation}
\label{mass-ex}
m^2_\phi(\sigma) =m_0^2\left(1+\alpha \frac{\sigma^2}{\Lambda^2}\right),
\end{equation}
where $\sigma$ is the light field and $\Lambda$ is the cut-off scale of
the effective action.
Note that a conventional interaction $\sim \alpha m_0^2 \sigma^2
\phi^2/\Lambda^2$ in the effective low-energy action may induce the
$\sigma$-dependent mass.
In this case, the thermal correction to the mass of the light field 
is $m_\sigma^2(T)\simeq \alpha(m_0^2/\Lambda^2)T^2$, which is supposed
to be smaller than the Hubble parameter $H^2\simeq Max\{\rho^{rad},
V_0\}/3M_p^2$, as in the inhomogeneous reheating scenario discussed in
Ref.\cite{Dvali-Mass-dom}. 
If there is no significant potential other than the finite-temperature
effective potential $V(\phi)_T$, we find an effectively flat $\sigma$
potential during the symmetry restoration phase.  
Since the interaction depends on the values of the fields $\sigma$ and 
$\phi$, the background field trajectories after the phase transition may
be sensitive to the initial conditions and the non-perturbative effects of
the decay process, which means that the general evaluation of the
cosmological parameters after the phase transition typically requires
numerical calculations \cite{PR-original}. 
However, the numerical study related to such a non-perturbative process
after the phase transition is highly model-dependent and out of the
scope of this paper. 
We thus assume that all the energy stored in the
potential is instantly converted into radiation just after the phase
transition, in order to single out the contribution from the
inhomogeneous phase transition. 
In addition to the complexities of the
decay process, the domain walls related to discrete
symmetry breaking may cause a problem. 
However, cosmological domain walls can be made
unstable and safe if a bias between the two vacua is induced by an
effective interaction term that breaks the $Z_2$ symmetry. 
Note that for supergravity, domain walls caused by R-symmetry are safe
since the supergravity interaction creates the required bias
\cite{matsuda-dwall}. 
Therefore, for simplicity and to allow calculation of the
model-independent contribution from the inhomogeneous phase transition,
we ignore the domain wall problem in this paper, expecting that the
walls decay instantly into radiation due to the bias between the two
vacua.  

\begin{figure}[h]
 \begin{center}
\begin{picture}(400,300)(0,0)
\resizebox{12cm}{!}{\includegraphics{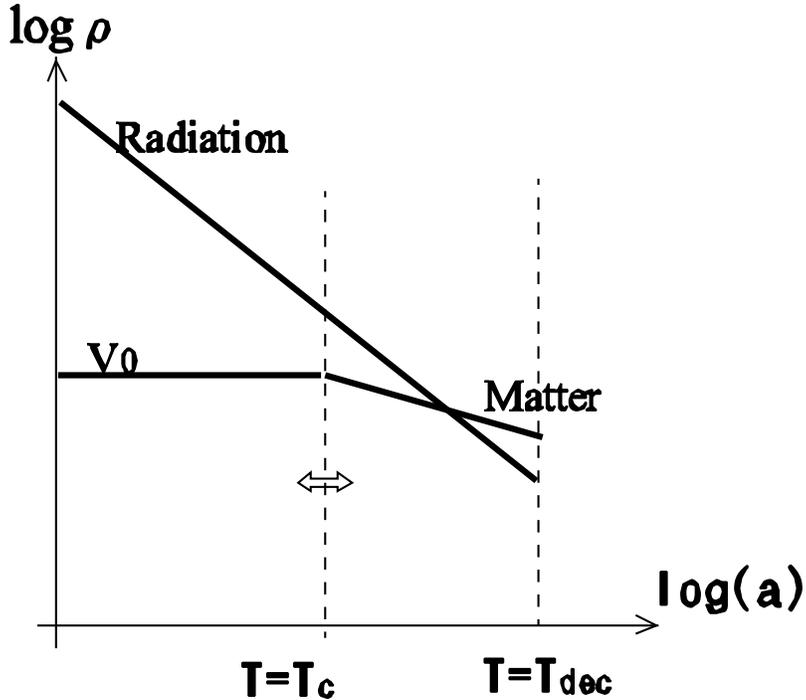}} 
\end{picture}
\caption{Initially, the Universe is dominated by radiation. The
  potential energy is converted into non-relativistic matter at $T=T_c$.
  Then the matter decays into radiation at $T=T_{dec}$. 
In the picture we show a case in which the non-relativistic matter
  dominates before the decay, but it is possible to consider a case in
  which the decay into radiation occurs before domination, as is
  discussed in the text.} 
\label{Fig:second-ex}
 \end{center}
\end{figure}

The latter scenario is less trivial. 
Let us consider the case in which the energy density of the Universe is
dominated by radiation at $T=T_c$ and 
the non-relativistic $\psi$ particles decay into radiation at
$T=T_{dec}<T_c$, as is shown in Fig. \ref{Fig:second-ex}.
We assume that all the potential energy is translated into $\psi$
particles at $T=T_c$.
We introduce the ratio $r\equiv\rho_\psi/\rho$ and consider the
case in which $\psi$ does not dominate 
the universe (i.e., $r(t_{dec})<1$).
Assuming that the symmetry restoration phase starts at some uniform
temperature $T=T_R$, and introducing $\rho^r_{vh}$ as before, 
at $t=t_{dec}$ we find
\begin{equation}
\rho_\psi(t)\propto \left(\frac{T_{R}}{T_c}\right)^4
\left(\frac{T_{c}}{T_{dec}}\right)\rho^r_{vh},
\end{equation}
where $\rho^r_{vh}$ scales like radiation.
The decay temperature $T_{dec}\equiv T(t_{dec})$ is determined by 
\begin{equation}
\rho(t_{dec}) \simeq \Gamma_\psi^2M_p^2,
\end{equation}
where we assume $\delta \Gamma_\psi=0$.
Therefore, the density perturbation is given by
\begin{equation}
\frac{\delta \rho}{\rho}=-3r\frac{\delta T_c}{T_c}.
\end{equation}

It is possible to consider a case in which the potential
energy decays into radiation
immediately after the phase transition (i.e., for $r(t_c)<1$ and
$T_c=T_{dec}$).
In this case, we find 
\begin{equation}
\rho_\psi(t)\propto \left(\frac{T_{R}}{T_c}\right)^4
\rho^r_{vh},
\end{equation}
which leads to the density perturbation given by
\begin{equation}
\frac{\delta \rho}{\rho}=-4r\frac{\delta T_c}{T_c}.
\end{equation}
\subsection{Non-thermal trapping}
In the simple second-order example, we consider effective couplings that
depend on light fields. 
Long-wavelength inhomogeneities of the light fields may lead to an
inhomogeneous critical temperature $\delta T_c\ne 0$.
Here, we consider another example, in which long-wavelength
inhomogeneities of the number density of some particles cause an
inhomogeneous end of the symmetry restoration phase.  

During preheating, some of the kinetic energy of the inflaton is
converted into excitations of the preheat field $\chi$. 
If $\chi$ couples to a field $\phi_a$ with a potential  
\begin{equation}
V(\phi_a,\chi)\sim V_0 -\frac{1}{2}m^2 \phi_a^2 
+\lambda \frac{\phi_a^{n}}{\Lambda^{n-4}}-\frac{g^2}{2}\phi_a^2\chi^2,
\end{equation}
where the inflaton terms are omitted, the effective potential caused by
the high density of the preheat field is given by\cite{End-trap, beauy-is}
\begin{equation}
V^{eff}(\phi_a)\simeq V_0-\frac{1}{2}m^2 \phi_a^2 
+\lambda \frac{\phi_a^{n}}{\Lambda^{n-4}} +g|\phi_a| n_\chi.
\end{equation}
For $\phi_a>0$, the effective potential gives
\begin{equation}
V^{eff}(\phi_a)\simeq V_0-\frac{1}{2}m^2 
\left(\phi_a^2 -\frac{gn_\chi}{m^2}\right)^2+
\frac{g^2 n_\chi^2}{2m^2}.
\end{equation}
When $n_\chi$ is very large, the field $\phi_a$ is
trapped by a strong attraction from the origin.
During the interval of the trapping, the potential barrier decreases,
since $n_\chi$ scales as $n_\chi\propto a^{-3}$, and ultimately
tunneling occurs below the critical number density \cite{End-trap,
beauy-is} given by
\begin{equation}
n_\chi \le n_c\equiv \frac{m^3}{g}.
\end{equation}
If the energy density is dominated by the potential energy $V_0$, the
trapping leads to an inflationary expansion. 
The number of e-foldings
elapsed during this interval is 
\begin{equation}
N =\frac{1}{3}\ln\left(\frac{n_\chi(t_i)}{n_\chi(t_e)}\right),
\end{equation}
where $t_i$ and $t_e$ are the time when the domination by the potential
energy starts and when the inflationary expansion ends.
In this model, inhomogeneities in the initial number density 
$n_\chi(t_i)$ can be created by inhomogeneous preheating\cite{IH-PR,
IH-PR-withoutID}.\footnote{Inhomogeneous preheating accompanied by
instant decay may directly lead to the creation of curvature
perturbation\cite{IH-PR}. In this section, we consider a preheating field
that does not lead to instant decay\cite{beauy-is,
IH-PR-withoutID}.}
The inhomogeneities in the preheating arise from the long-wavelength
fluctuations of the multi-field trajectory for the symmetry-breaking
potential. 
This is the origin of $\delta N$ discussed in Ref.\cite{End-trap}.
In addition to the inhomogeneities in $\delta n_\chi(t_i)$, we
may consider inhomogeneities in $n_c$, which are non-zero if $m$ and
$g$ are modulated at the end of the trapping phase.
Using the delta-N formalism, for $\delta n_\chi(t_e)\simeq
\delta n_c$ we find
\begin{equation}
\delta N =-\frac{1}{3}\frac{\delta n_c}{n_c}\simeq
-\frac{\delta m}{m}+\frac{1}{3}\frac{\delta g}{g}.
\end{equation}

\subsection{Electroweak phase transition}
The main obstacle in building a model of an inhomogeneous electroweak
phase transition is that the long-wavelength inhomogeneities of the
light field must survive until the electroweak phase transition, when
the Hubble parameter is much lower than the gravitino mass. 
In supergravity models inspired by string theory, there are many light
fields (moduli) in the effective action,
 but typically the mass of the
moduli fields is expected to be of the same order as the gravitino mass,
where the gravitino mass is generically given by 
$m_{3/2}\sim \Lambda_{SUSY}^2/M_p$, where $\Lambda_{SUSY}>$ TeV is
the supersymmetry breaking scale\cite{matsuda_moduli}.
The mass of the moduli $\sim m_{3/2}$ is clearly larger than 
$H_{EW}\equiv T_{EW}^2/M_p$, where $T_{EW}$ is the critical temperature
for the electroweak phase transition.
Therefore, if an inhomogeneous phase transition occurs at the
electroweak phase transition, the inhomogeneities of the effective
action must be inherited from the fluctuations of the light fields whose
potential is protected by symmetry before the electroweak phase
transition, while the moduli potential must be lifted after the
phase transition. 
The above condition for the inhomogeneous electroweak phase transition
might seem very severe, but in string theory there is at least one
specific example that may induce an inhomogeneous phase transition at
the electroweak scale. 
We consider intersecting D-brane models, which
are an interesting possibility for string model building, allowing us to
devise models that are sensibly close to the Minimal Supersymmetric
Standard Model (MSSM) in terms of particles and gauge groups
\cite{Int-Brane-Review}. 
A remarkable feature of this scenario is that the flavor structure of the
Yukawa couplings may arise from the matter fields located at different
intersections, with the resulting Yukawa couplings expressed by the
classical instanton action of the minimal world-sheet area: 
\begin{equation}
Y\propto \exp\left(-\frac{A}{2\pi \alpha'}\right),
\end{equation}
where $A$ is the minimal world-sheet area of the intersection.
If the model is constructed from D6-branes in Type IIA string theory
wrapping orientifolds of $R_4\times T^2\times
T^2\times T^2$\cite{Int-Brane-Review}, there will be shift symmetries
that correspond to the brane motion in the internal space.
If the shift symmetries are not broken in the effective action, the
minimal world-sheet area remains as an arbitrary parameter. 
Considering moduli fields $\sigma_i$, $(i=1,2,3)$ for the three branes
constituting a triangle in the internal space, we find;
\begin{equation}
\delta A(\sigma_i)
\simeq \sum_{i} \frac{\partial A}{\partial \sigma_i} \delta \sigma_i + 
\sum_{ij} \frac{\partial^2 A}{\partial \sigma_i\partial \sigma_j} 
\delta \sigma_i \delta \sigma_j.
\end{equation}
It would be better for our purpose to consider a simple form of
$A(\sigma)$ and consider the inhomogeneity $\delta \sigma$ to obtain
\begin{equation}
\delta A(\sigma)
\simeq A' \delta \sigma + A'' (\delta \sigma)^2
\equiv \alpha_1 \frac{\delta \sigma}{\Lambda_A}+
\alpha_2 \left(\frac{\delta \sigma}{\Lambda_A}\right)^2.
\end{equation}
Let us consider a possible mechanism for generating an effective
potential related to $A$. 
Assuming that the Yukawa couplings are generated by the mechanism and
considering the standard one-loop correction to the Higgs field
potential from the top fermion loop, we obtain for 
\begin{equation}
\label{effective}
\Delta m_H^2 \sim -\frac{3}{4\pi^2} Y_t^2 m_{\phi_t}^2\ln\left(
\frac{\mu}{ m_{\phi_t}}\right),
\end{equation}
where $m_{\phi_t}$ denotes the scalar top mass\cite{EW-DSB}.
Here we consider $Y_t \propto \exp(-A/(2\pi \alpha'))$.
In the MSSM, the one-loop correction from the top Yukawa coupling
destabilizes the Higgs potential and causes electroweak symmetry
breaking. 
From the one-loop correction term in Eq. (\ref{effective}), we find that
the world-sheet area can be stabilized after the electroweak symmetry
breaking\cite{Matsuda_yukawa}. 
In this case, the free motion of the D6-branes in the internal space is
protected by the shift symmetry before the electroweak symmetry
breaking. 
However, after the electroweak 
symmetry breaking, which is induced by the loop correction in the MSSM
electroweak symmetry-breaking scenario, the shift symmetries are partly
broken and the minimal world-sheet area is stabilized in the low-energy
effective action.  
Although the scenario
depends greatly on the specific details of the intersecting brane
models, a generic implication of the scenario is that the inhomogeneous
phase transition occurs whenever the shift symmetries are not explicitly
broken before the phase transition. 
A similar mechanism may work at the
GUT phase transition, and the inhomogeneous phase transition may lead to
a cosmological signature of the intersecting brane models. 

\section{Conclusions and discussions}
We have studied a mechanism for generating primordial density
perturbations in inflationary models. 
We considered long-wavelength
inhomogeneities of light scalar fields that cause superhorizon
fluctuations of couplings and masses in the effective low-energy
action. 
Since the effective couplings and masses are not homogeneous in
space, cosmological phase transitions may occur that are not
simultaneous in space. 
It is possible to create the primordial curvature
perturbation from the mechanism, but more generally the scenario of an
inhomogeneous phase transition allows for non-Gaussianity to occur in
the spectrum after inflation \cite{Bartolo-text, NG-obs}. 
It is useful to specify the level of non-Gaussianity by the non-linear
parameter $f_{NL}$, which is usually 
defined by the Bardeen potential $\Phi$,
\begin{equation}
\Phi=\Phi_{Gaussian}+f_{NL}\Phi_{Gaussian}^2.
\end{equation}
Using the Bardeen potential, the curvature perturbation $\zeta$ is given
by
\begin{equation}
\Phi=\frac{3}{5}\zeta.
\end{equation}
When we consider ``additional'' non-Gaussianity created at the
inhomogeneous phase transition, the first-order perturbation is
generated dominantly by the usual inflaton perturbation. 
Therefore, the ``additional'' second-order perturbation is not
correlated to the first-order perturbation. 
In this case, the non-linear parameter is estimated as \cite{unco-NG}
\begin{equation}
\frac{6}{5}f_{NL}\simeq \frac{1}{N_\phi^4}\left[
N_\sigma^2N_{\sigma\sigma}+N^3_{\sigma\sigma}{\cal P}_\sigma
\log(k_b L)
\right],
\end{equation}
where $\zeta$ can be expanded by the $\delta N$ formalism as
\begin{equation}
\zeta \simeq N_\phi \delta \phi + N_\sigma\delta\sigma
+\frac{1}{2} N_{\phi\phi} \delta \phi^2 
+\frac{1}{2} N_{\sigma\sigma} \delta \sigma^2 + ...,
\end{equation}
and we assume that the perturbation can be separated as
\begin{equation}
\zeta = \zeta^{(\phi)}+\zeta^{(\sigma)}.
\end{equation}
Here $k_b\equiv$ min$\{k_i\}$ $(i=1,2,3)$ is the minimum wavevector of the
bispectrum and $L$ is the size of a box in which the perturbation is
defined.
A useful simplification is\cite{Lyth_and_Rod_NG} 
\begin{equation}
f_{NL}\simeq \left(\frac{1}{1300}
\frac{N_{\sigma\sigma}}{N_\phi^2}\right)^3.
\end{equation}
The scenario of adding non-Gaussianity from the inhomogeneous phase
transition is interesting, since for the effective low-energy action,
higher-dimensional couplings may naturally appear with light fields in a
decoupled sector. 

Consider a simple example discussed in Sec. 2.1 with $\delta m_\phi\ne
0$ and $\delta g =\delta \lambda =0$.
Considering the initial value for the light field $\sigma$ in
Eq.(\ref{mass-ex}), a modest assumption would be $\sigma \simeq 0$.
From Eq. (\ref{delm}), the curvature perturbation created from the
inhomogeneous phase transition is purely second order and given by 
\begin{equation}
\zeta^{(\sigma)} \simeq -\frac{\delta m_\phi}{m_\phi} 
\simeq -\alpha
\frac{(\delta \sigma^2)}{2\Lambda^2}
= -
\frac{\alpha H_I^2}{2\Lambda^2(2\pi)^2},
\end{equation}
 where $H_I$ is the Hubble parameter when the
long-wavelength inhomogeneity of the light field $\sigma$ exits the
horizon during inflation. 
Thus we find from the $\delta N$ formula;
\begin{equation}
N_{\sigma\sigma} = -
\frac{\alpha}{\Lambda^2}.
\end{equation}
Even for the initial condition $\sigma \simeq 0$, the non-linear
parameter for the inhomogeneous phase transition is significant.
Considering the usual normalization for the first order perturbation,
we find\cite{EU-book}
\begin{equation}
|N_\phi \delta \phi| \simeq 5 \times 10^{-5}.
\end{equation}
The non-linear parameter is thus given by
\begin{equation}
f_{NL} \simeq \left(10^{6} \times \alpha 
\frac{H_I^2}{\Lambda^2}\right)^3.
\end{equation}
Considering the modest bound for the non-linear parameter $|f_{NL}|<100$,
the above result puts a significant upper bound on the
inflationary scale or on the effective couplings that contain decoupled
light fields. 

For the electroweak phase transition, we find for the simple case
($A\equiv A(\sigma)$); 
\begin{equation}
\zeta^{(\sigma)} = -\frac{\delta (\Delta m_H)}{\Delta m_H} 
\simeq \frac{\delta A}{2\pi \alpha'}
\simeq  \frac{\alpha_1}{2\pi \alpha'} \frac{\delta \sigma}{\Lambda_A}+
\frac{\alpha_2}{2\pi \alpha'} 
\left(\frac{\delta \sigma}{\Lambda_A}\right)^2.
\end{equation}
With regard to the non-Gaussianity, we find from the above equation
that $\alpha$ in the standard calculation is simply replaced by
$-\alpha_2/(2\pi \alpha')$ for the electroweak phase transition with
the effective scale $\Lambda_A = \Lambda$.

Inhomogeneous phase transition can be applied to warm inflationary
scenario. 
Inhomogeneous phase transition in warm inflation will be discussed in
our forthcoming paper\cite{remote-matsuda}.

\section{Acknowledgment}
We wish to thank K.Shima for encouragement, and our colleagues at
Tokyo University for their kind hospitality.
\appendix
\section{$\delta N$ formalism for the curvature perturbation}

Here we consider two different definitions for the curvature
perturbations.
The comoving curvature perturbation (${\cal R}$) can be related to
 the curvature perturbation on uniform-density hypersurfaces ($\zeta$)
 by studying the evolution at large scales.
The gauge-invariant combinations for the curvature
perturbations can be constructed as follows:
\begin{eqnarray}
\label{zeta-org}
\zeta &=&-\psi -H\frac{\delta \rho}{\dot{\rho}}\nonumber\\
{\cal R}&=& \psi -H\frac{\delta q}{\rho+p},
\end{eqnarray}
where $\delta q$ is the momentum perturbation that is expressed as
$\delta q =-\dot{\phi}\delta \phi$ for the inflaton $\phi$ with a standard
kinetic term.
Linear scalar perturbations of a
Friedman-Robertson-Walker(FRW) background were considered:
\begin{equation}
ds^2=-(1+2A)dt^2 + 2a^2(t)\nabla_i B dx^i dt +a^2(t)
[(1-2\psi)\gamma_{ij}+2\nabla_i\nabla_j E]dx^i dx^j.
\end{equation}
Here $\rho$ and $p$ denote the energy density and the pressure.
Spatially flat hypersurfaces and uniform density hypersurfaces
are defined by $\psi=0$ and $\delta \rho=0$, respectively.\footnote{In
this Appendix, ``$\psi$'' is used for a metric perturbation.} 

Besides the curvature perturbations defined above, it is useful to
define the perturbed expansion rate with respect to the 
coordinate time.
The perturbed expansion rate is expressed as
\begin{equation}
\delta \tilde{\theta}\equiv -3\dot{\psi}+\nabla^2\sigma,
\end{equation}
where the scalar describing the shear is
\begin{equation}
\sigma = \dot{E}-B.
\end{equation}
Choosing the gauge whose slicing is flat at $t_{ini}$ and uniform
density at $t$, the $\delta N$ formula is given by
\begin{equation}
\zeta = \frac{1}{3}\int^t_{t_{ini}}\delta \tilde{\theta}dt =\delta N.
\end{equation}
The $\delta N$ formula is sometimes expressed by
\begin{equation}
\zeta=\delta N=-H \left.\frac{\delta \rho}{\dot{\rho}}\right|_{\psi=0},
\end{equation}
where $\delta N$ is the perturbed expansion to uniform-density
hypersurfaces with respect to spatially flat hypersurfaces,
and $\delta \rho$ must be evaluated on spatially flat hypersurfaces.

\end{document}